\documentclass{svproc}
\usepackage{url}

\usepackage{xspace}
\newcommand{\method}{\textsc{SCG}\xspace}
\newcommand{\methodmap}{\method-map\xspace} 
\newcommand{\methodflag}{\method-flag\xspace} 
\newcommand{\methodsig}{\method-signature\xspace} 
\newcommand{\methodmeso}{\method-meso\xspace} 
\newcommand{\dense}{coordinated\xspace}

\newcommand{\scale}{{Scalability\xspace}}

\newcommand{\content}{{Content\xspace}}
\newcommand{\connection}{{Connection\xspace}}
\newcommand{\tinyc}{{Tiny Cluster\xspace}}
\newcommand{\character}{{Characterization\xspace}}

\newcommand{\bit}{\begin{compactitem}}
\newcommand{\eit}{\end{compactitem}}
\newcommand{\ben}{\begin{compactenum}}
\newcommand{\een}{\end{compactenum}}

\newcommand{\ie}{i.e.\xspace}

\usepackage{graphicx}

\usepackage{diagbox}

\usepackage{dingbat}

\usepackage{multirow}

\usepackage{stmaryrd}

\usepackage{amsfonts}

\usepackage{amsmath}

\begin{document}
\mainmatter              
\title{\textit{\method}: Spotting Coordinated Groups in Social Media}
\titlerunning{\method}  
%
\author{Junhao Wang\inst{1,2} \and Sacha Levy\inst{1,2}
Ren Wang\inst{3} \and Aayushi Kulshrestha\inst{1,2} \and Guillaume Rabusseau\inst{4,2} \and Reihaneh Rabbany\inst{1,2}}
\authorrunning{Junhao Wang et al.} 
%
%
\institute{McGill University, Montreal, Canada
\and
Mila, Montreal, Canada
\and UBC, Vancouver, Canada
\and Université de Montréal, Montreal, Canada}

\maketitle              

\begin{abstract}
Recent events have led to a burgeoning awareness on the misuse of social media sites to affect political events, sway public opinion, and confuse the voters.
Such serious, hostile mass manipulation has motivated a large body of works on bots/troll detection and fake news detection, which mostly focus on classifying at the user level based on the content generated by the users. 
In this study, we jointly analyze the connections among the users, as well as the content generated by them to Spot Coordinated Groups (\method), sets of users that are likely to be organized towards impacting the general discourse. Given their tiny size (relative to the whole data), detecting these groups is computationally hard. Our proposed method detects these tiny-clusters effectively and efficiently. 
We deploy our \method method to summarize and explain the coordinated groups on Twitter around the 2019 Canadian Federal Elections, by analyzing over 60 thousand user accounts with 3.4 million followership connections, and 1.3 million unique hashtags in the content of their tweets. 
The users in the detected coordinated groups are over 4x more likely to get suspended, whereas the hashtags which characterize their creed are linked to misinformation campaigns.

\keywords{social network, misinformation, graph mining}
\end{abstract}

\section{Introduction}

\begin{figure*}[t]
\centering
  \includegraphics[width=.97\linewidth]{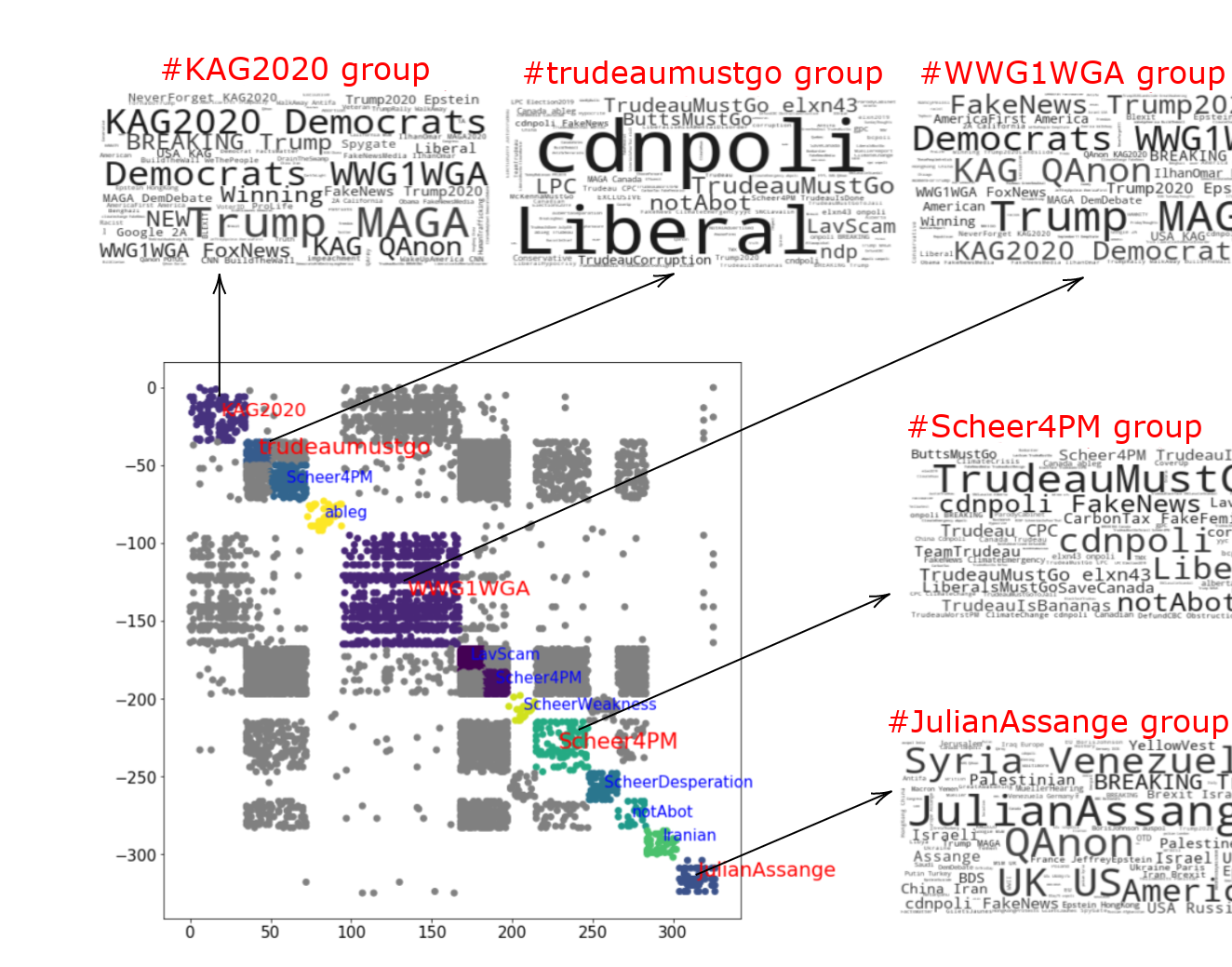}
  \caption{\method detects 13 \dense groups in the 2019 Canadian Federal Election including multiple \#MAGA groups. }
  \label{fig:blockssus}
  \vspace{-5pt}
\end{figure*}

Manipulation of online discourse through social media is a pressing global concern \cite{starbird2019disinformation}. 
Recently, the Special Counsel for the U.S. Department of Justice published their investigation into Russian "Active Measures" social media campaign, which confirmed an \textit{organized} attempt at the state level to sow discord into the U.S. political system through social media \cite{mueller2019report}. 
As an example, Twitter reported possible engagement of 1.4 billion users with the suspected "trolls" from the Russian government funded Internet Research Agency (IRC) \cite{policy2018update}, and it is believed that this interference has swayed the 2016 US Presidential Election \cite{badawy2018analyzing}.
Such activities aim to distort information space to confuse and distract voters, disseminate propaganda and disinformation to foster divisions, and  paralyze the decision making abilities of individuals \cite{wilson2018assembling}. 
The ultimate goals or motives of these operations might be hard to interpret, but their effect on public opinion, democracy and elections is clear \cite{marwick2017media,bovet2019influence}. 

The severity and scale of such operations motivated the social media giants like Twitter and Facebook to update their site policies \cite{TwitterElect,FbInauth}. 
These updated policies aim at tackling \textit{Information Operations} - the suit of methods used to influence others through the dissemination of propaganda and disinformation \cite{TwitterElect,wilson2018assembling}, and \textit{Coordinated Inauthentic Activities} - groups working together to mislead people about who they are and what they are doing \cite{FbInauth}. 

How can we monitor the information space proactively, identify such coordinated activities at an early stage, to ensure a healthy democratic society? 
In this paper we present \method as a solution to this problem. \method is a novel framework that consists of modular components to study the activity in complex social media space and identify groups that are impacting the information space in an organized manner.
In extreme cases, bots generated from the same script behave in an almost identical or highly correlated manner \cite{chavoshi2016identifying}, and trolls or sockpuppets, who are being operated by the same person behind the scenes, exhibit lock-step behavior \cite{kumar2017army}. 
More generally, \method detects {\dense groups}, members of which \textit{amplify each others' voice and boost each others' influence}, unlike what is the norm among typical users. These are suspicious groups that need to be \textit{further investigated by a human} as detecting trolls is not a trivial task \cite{link1} and political campaigns or activist groups might exhibit similar coordinated behaviour. 
\method makes this subsequent investigation efficient by providing a group level summarization and characterization.

We formulate this problem as finding { tiny clusters} (relative to the size of online society)  which introduce \textit{dense regions on coupled matrices of users' activity (content) and users' interactions (connection)}. 
Through controlled synthetic experiments, we demonstrate the difficulty of this regime for different off-the-shelf contenders and the \textit{superior performance} of our proposed \method in terms of both \underline{scalability} and \underline{accuracy} in recovering the injected tiny clusters.

\begin{figure*}[t]
\centering
  \includegraphics[width=.97\linewidth,height=.45\linewidth,trim=2.8in 0.2in 0in 0in, clip]{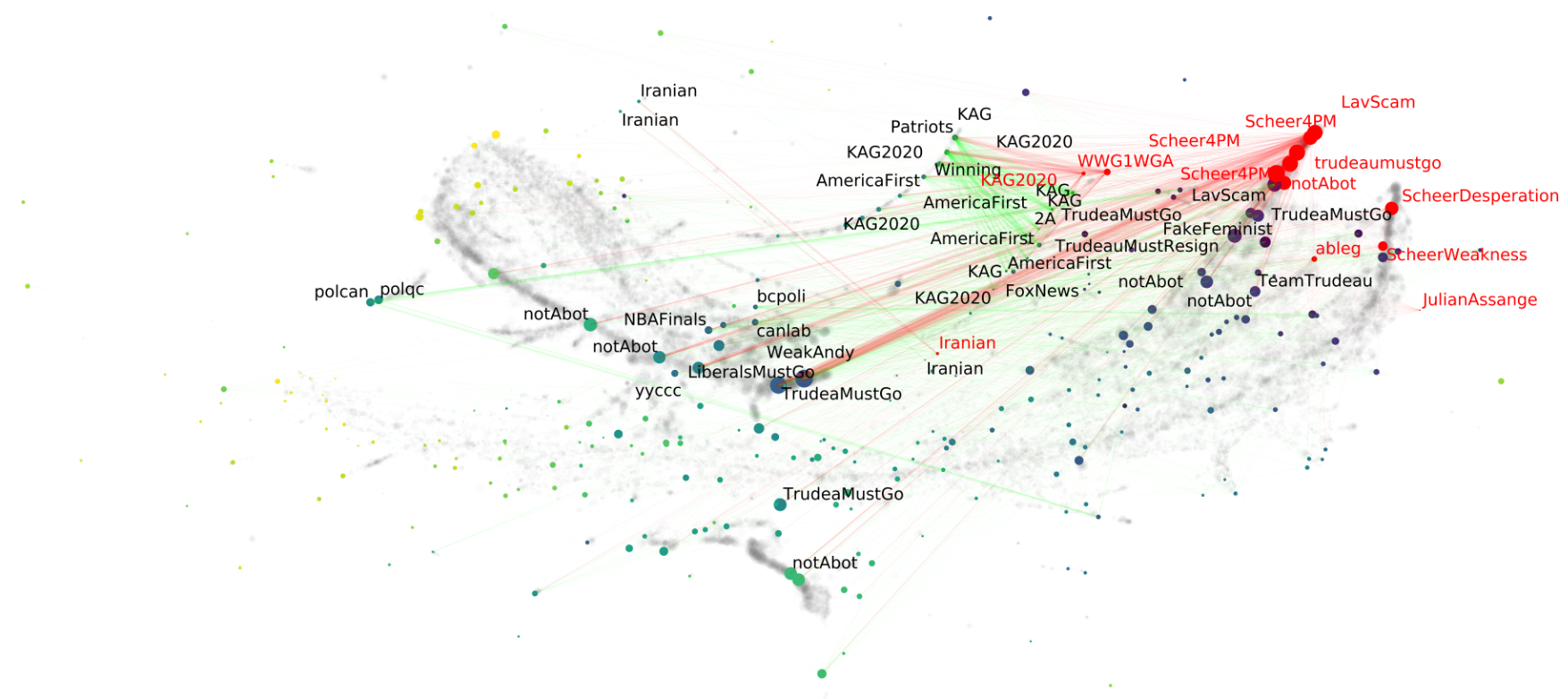}

\caption{\method summarizes Twitter dynamics of {69,709 accounts} around 2019 Canadian Election, providing a bird's eye view of how detected \dense groups (marked red) engage in the overall discourse.}
  \label{fig:mesotop}
  \vspace{-10px}
\end{figure*} 
 
We employ \method to spot \dense groups active around the 2019 federal elections in Canada. In particular, we monitor the activity within Twitter, the most commonly used platform to mobilize the public at the time of political unrest. Figure \ref{fig:blockssus} shows the 13 \dense groups flagged, and Figure \ref{fig:mesotop} maps out how these groups engage in the overall discourse. 
A key observation in Figure \ref{fig:blockssus} is a  block of multiple conservative groups that engage with both American conservative politics and Canadian election politics. This block encompasses Trump \#MAGA US groups as well as a domestic anti-Trudeau group. 
The different popularity of hashtags used within each group reflects their different political creed. The grey background shows the connections among these \dense groups. %
User accounts in these detected \textit{\dense groups are more than four times more likely to be suspended} by Twitter.
On the other end, several of the hashtags these \dense groups use, singled out by \method, \textit{are linked to the misinformation campaigns}, in particular \#notAbot and \#TruedeauMustGo  \cite{emma19,Orr19}.
While these observations are important and help us better navigate the map of twitter activity around this election, we want to emphasize that \textit{\method provides a} \textit{general and novel tool }for computational social science and could be applied to study and investigate suspiciously organized groups engaged around any given topic in any online social network. The main contributions of our work can be further  summarized with respect to \method's modules, i.e.:

\begin{itemize}
    \item \methodmap fuses user \textbf{connections} and the \textbf{content} they post to embed large scale activity on social media in a holistic fashion
    \item  \methodflag \textbf{effectively} and \textbf{efficiently} detects \textbf{tiny clusters} of users who are posting similar content and are densely connected 
    \item  \methodsig and \methodmeso \textbf{characterize} and explain the group creeds and how they engage with each other and the rest of users
     \item   \textbf{Scalability}
     : \method scales \textit{linearly} with the number of edges in the graph 
\end{itemize}

\textbf{Deployment}: our project was deployed to monitor the 2019 Canadian Federal Elections, the main election in Canada which indirectly elects the prime minister (head of the majority party). The data collection process started in April 2019, and the analysis process started in September 2019, and results were first published October 2019 right before the election \cite{wang2019sgp}. The findings were reported before the election \cite{link2,link3}, and were corroborated by other independent research groups and investigators afterwards \cite{link4,rheault2020investigating}. We have released an interactive dashboard for better interpretation of our results and started a 2-years close collaboration with political scientists as the followup of this study to understand the implications of the findings, deploy the improved system on future upcoming elections, and  be politically smart in the age of misinformation \cite{link5}. 

\textbf{Reproducibility}: the supplementary materials of \method code (including code to generate the synthetic data), extended version of the paper including appendices that explain the details of method and experiments, as well as the visualization dashboard to investigate the results are released\footnote{\url{https://sites.google.com/view/spg-exp}}. The details of data collection is also  explained in the supplementary materials, however Twitter policies do not allow re-sharing the crawled data.

\section{Background and Related Work}

In this section, we briefly review the related works on online misinformation detection and general dense block detection methods. 

\subsection{Detecting Misinformation Online}

Many recent works \cite{badawy2018analyzing,mitchell_gottfried_kiley_matsa_2014,wilson2018assembling} analyze the vulnerabilities of social media to information operations and coordinated inauthentic activities, and relate them to the clustering of politicized online information spaces. This phenomenon, defined as "echo-chambers", describes the gathering of like-minded individuals on online communities. 
As illustrated by Marwick \textit{et al} \cite{marwick2017media}, the defiance toward traditional media from part of the population leads to the emergence of alternative (possibly biased or fake) news sources. 
Bovet \textit{et al.} \cite{bovet2019influence} showed that Twitter trolls tend to form small, politically biased groups that propagate misleading information to normal users. Stewart \textit{et al} \cite{stewart2018examining} identified trolls as polarising elements of echo-chambers, distorting the information space. 
Most past works on online misinformation detection are largely limited to classifying  user-generated content or users, based on their activities \cite{shu2017fake,zhou2018fake,zhou2019fake,shu2019defend}. 
Unlike these supervised techniques, we take a novel unsupervised approach that jointly analyzes content and user connections, inspired by successful application of unsupervised  techniques in anomaly and fraud detection settings, for example, to detect fake reviewers posted to artificially boost product ratings on E-commerce sites \cite{DBLP:conf/kdd/HooiSBSSF16}.

\begin{table}[t]
\small
\begin{center}
\begin{tabular}{ l| c  c |c  c  c | c  c ||c}
      
      \diagbox{Property}{Method}   & \rotatebox{90}{Louvain\cite{blondel2008fast}} &   \rotatebox{90}{Infomap\cite{rosvall2008maps}} &
       \rotatebox{90}{node2vec\cite{grover2016node2vec} } &
       \rotatebox{90}{attri2vec \cite{zhang2019attributed}} &
       \rotatebox{90}{GraphSAGE\cite{hamilton2017inductive}} &
       \rotatebox{90}{pcv \cite{NIPS2018_7643}} &
       \rotatebox{90}{Fraudar\cite{DBLP:conf/kdd/HooiSBSSF16}} & \rotatebox{80}{\method} \\ 
\hline  
        \content & & & &\checkmark &\checkmark &\checkmark & &\checkmark \\
        
        \connection &\checkmark &\checkmark &\checkmark &\checkmark &\checkmark & &\checkmark &\checkmark\\ 
        
	    \tinyc &  & & & & &\checkmark &\checkmark &\checkmark\\ 
	    
	     \character & & & & & & & &\checkmark  \\
	    
        \scale &  &\checkmark & &  & &\checkmark &\checkmark &\checkmark
        \\
\end{tabular} 
\caption{ {\bf \method matches all specs}, while competitors
miss one or more of the desired features.\label{tab:salesman}}
\end{center}
\vspace{-15pt}
\end{table}

\subsection{Dense Block Detection}
Given the  lockstep behavior exhibited by anomalous users, multiple dense block detection methods have been designed for anomaly and fraud detection \cite{shin2016m,shin2016corescope,shin2017d}.
Finding exact cliques and quasi-cliques is shown to be NP-hard \cite{lee2010survey}. Classical solutions for the clique problem can be categorized as exact enumerations \cite{regneri2007finding}, fast heuristic enumerations \cite{kumar1999trawling} and bounded approximation algorithms \cite{charikar2000greedy}, most of which have runtime at least polynomial to the size of graph. 
Fraudar\cite{DBLP:conf/kdd/HooiSBSSF16} is a notable example of scalable dense subgraph detection methods that finds subgraphs with large average degree in the context of fraud/anomaly detection.  
Various extensions of dense subgraph detection include dense sub-tensor detection \cite{shin2016m,shin2017d}, online dense sub-tensor detection \cite{shin2017densealert}, hierarchical dense subgraph detection \cite{zhang2017hidden}. These methods are defined in a single mode whereas our method detects coupled blocks which enforce dense substructures in coupled matrices/graphs as discussed later in detail. 

\subsection{Graph Clustering}
Dense subgraph detection is closely related to graph clustering or community detection, which identifies clusters of densely connected nodes \cite{yang2011detecting,kelley2012defining}. Two widely-used community detection algorithms include Louvain \cite{blondel2008fast}, based on modularity optimization and Infomap \cite{rosvall2008maps}, based on information compression. Table \ref{tab:salesman} puts these methods in contrast with our proposed \method. 

On the other end, given a node embedding method, we can run any clustering algorithm to recover network community structure. Network embedding techniques aim to map nodes or subgraphs onto Euclidean space through possibly learned functions on graphs. The most notable graph embedding techniques are unsupervised GraphSAGE \cite{hamilton2017inductive}, node2vec \cite{grover2016node2vec} and attri2vec \cite{zhang2019attributed}. We show in our experiments that these methods fail to recover tiny clusters effectively.  
\subsection{Finding Tiny Clusters}
It is well-known that modularity optimization fails to identify clusters smaller than a scale \cite{fortunato2007resolution}. This resolution limit depends on the size of the network and the interconnectedness of the clusters. Few works try to discover clusters of small size in graphs \cite{xu2014jointly,lim2015convex}. Notably, pcv method  \cite{NIPS2018_7643} considers bipartite stochastic block models and finds tiny clusters with theoretical guarantees. Our work takes a similar notion of tiny cluster, but in a more general case of coupled matrices as explained later. This combining of the different sources of information is proven to be a necessity for better recovery of community structure 
\cite{deshpande2018contextual}.

\section{Proposed Method}

How can we find a \textit{\dense group} given millions of records of users' activities in online social platforms?
A group that consists of a \textit{relatively small set of accounts that aim to increase their influence through following each other and broadcasting similar  messages.}
Looking at the connections among the user accounts, a \dense groups forms subgraph of much higher density than the background. Similarly, dense subgraphs exits on the bipartite network of users in \dense groups and their messages/content they post. 
This can be captured by artifacts such as hashtags or noun-phrases from the content and can also be considered as attributes on the user nodes.
Furthermore, these \dense groups are much smaller in size than naturally formed network communities or clusters, and our interest is to recover exactly these \textit{tiny clusters}, instead of the global community structure of the network. More formally we define: 

\begin{definition}[\dense group] \label{def:pollutegroupnew}
Given user connection graph with adjacency matrix  $\mathbf{A}\in\{0,1\}^{n\times n}$ and user-attribute bipartite graph with biadjacency matrix $\mathbf{X} \in \{0,1\}^{n\times d}$, a \dense group is the set of user nodes 
that induce high density on both $\mathbf{A}$ and similar attribute usage pattern on $\mathbf{X}$.
\end{definition}

To detect these \dense groups, we propose \method procedure. 
We design \method to be a modular framework that consists of four components: \methodmap \textbf{maps out} large scale activity on social media by jointly embedding user connections and the content they post; \methodflag \textbf{detects} groups of users which are close in the joint embedding which indicates that they are posting similar content and are also densely connected to each other; \methodsig  \textbf{characterizes} the engagement of the \dense groups and finds their group creed; \methodmeso  \textbf{explains} how different \dense groups engage with each other and the rest of the population. 

More specifically, given user connection adjacency matrix $\mathbf{A}$ and user attribute matrix $\mathbf{X}$, \methodmap creates user embedding $\mathbf{Z}= \mathbf{A} \phi(\mathbf{X})$
where $\phi$ projects $\mathbf{X}$ on its first k left singular vectors through truncated singular value decomposition. 
\methodflag clusters $\mathbf{Z}$ using k-means and ranks the clusters by their induced density on $\mathbf{A}$. 
\methodsig creates an ordering for all attributes for each cluster, by computing the difference between in-cluster normalized usage frequency and global normalized usage frequency of each attribute. 
Finally, \methodmeso measures the strength of interaction between two user clusters as the amount of edges going though two user clusters, normalized by the product of their respective sizes. The exact formulas used and details of the implementation are provided in the extended version of the paper made available in the supplementary materials.

\section{Experiments}

In this section, we first verify the effectiveness and scalability of the first two components of the proposed \method method (\methodmap, \methodflag) through a set of synthetic experiments with builtin ground-truth, which approximate the real-world problem and enable us to provide a quantitative evaluation. 
Next, we discuss the observations provided by applying all four components of \method on scraped real-world Twitter data and provide several pieces of evidence on the effectiveness and interpretability of the \method in unveiling the dynamics of \dense groups around the 2019 Canadian federal election. We used user followership as user connection graph, and hashtag usage as user attribute graph. Due to time and budget constraints, we focus our analysis on data scraped from Twitter around this particular event. Incorporating other social media platforms and other events is part of the future works planned for this study. 

\subsection{Validation on Synthetic Data}

Based on observation of our scraped dataset, real-world graphs are large, sparse and have high-dimensional node attributes. To approximate real-world data that has ground truths for \dense groups, we generate synthetic attributed graphs with similar characteristics but with injected \dense groups that serve as the groundtruth\footnote{details and synthetic generation code are available in the supplementary materials}. We compare different methods on how well they are able to recover these injected \dense groups.

\paragraph{\textbf{Parameter settings:}} 
We generate eight \dense groups with 20 nodes and 20 attributes on differently sized graphs (2,000 to 30,000 nodes/attributes) to test the effectiveness and scalability of \method. This gradually decreases the ratio of coordinated nodes from 8\% to 0.5\% of the graph size, thus making the detection progressively more challenging. 

\paragraph{\textbf{Baselines:}} 
Literature on unsupervised detection of \dense groups is relatively sparse, thus we carefully select unsupervised baselines from related literature: Infomap \cite{rosvall2008maps} and Louvain \cite{blondel2008fast} from community detection; Fraudar \cite{DBLP:conf/kdd/HooiSBSSF16} and pcv \cite{NIPS2018_7643} from dense subgraph or tiny cluster detection; node2vec \cite{grover2016node2vec}, attri2vec \cite{zhang2019attributed} and unsupervised GraphSAGE \cite{hamilton2017inductive} from network embedding. As summarized in \method in Table \ref{tab:salesman}, pcv baseline only considers content, the Infomap, Louvain, Fraudar and node2vec only consider connections, and the other baselines incorporate both content and connections. For a subset of baselines (node2vec, graphSAGE, attri2vec), we only run them on graphs with size up to 18,000 nodes due to time and hardware constraints. 

\paragraph{\textbf{Evaluation:}}

To evaluate partitions (how well \dense groups are separated from the background and each other), we use Quality score in \cite{NIPS2018_7643}, given $k$ ground-truth clusters $U_{1\dots k}$ and $s$ inferred clusters $\widetilde{U}_{1\dots s}$, and $J(\cdot, \cdot)$ as the Jaccard similarity between two sets, the Quality score is:\vspace{-6pt}
\begin{equation}
    Q = \frac{1}{k} \sum_{i=1}^k \max_{j=1,\dots ,s} J(U_i, \widetilde{U}_j) \in [0,1]
\end{equation}

To evaluate the ability to classify nodes as belonging to a \dense group or not, we use the F1 score. We generate two instances of synthetic attributed graph for each size, and do two runs on each instance and report the mean performance across all four runs.

\paragraph{\textbf{Performance analysis:}}

\begin{table*}[]
\centering
\tiny
\begin{tabular}{c|cc|cc|cc|cc|cc|cc|cc|cc}
\hline
\hline
\multirow{2}{*}{} &
  \multicolumn{2}{c|}{$n=2000$} &
  \multicolumn{2}{c|}{$n=6000$} &
  \multicolumn{2}{c|}{$n=10000$} &
  \multicolumn{2}{c|}{$n=14000$} &
  \multicolumn{2}{c|}{$n=18000$} &
  \multicolumn{2}{c|}{$n=22000$} &
  \multicolumn{2}{c|}{$n=26000$} &
  \multicolumn{2}{c}{$n=30000$} \\
 &
  Quality &
  F1 &
  Quality &
  F1 &
  Quality &
  F1 &
  Quality &
  F1 &
  Quality &
  F1 &
  Quality &
  F1 &
  Quality &
  F1 &
  Quality &
  F1 \\ \hline
Louvain &
  3.92 &
  14.81 &
  2.70 &
  4.91 &
  2.43 &
  2.02 &
  2.29 &
  0.56 &
  2.33 &
  0.75 &
  2.12 &
  0.47 &
  1.97 &
  0.00 &
  2.07 &
  0.00 \\
Infomap &
  11.11 &
  0.00 &
  11.11 &
  0.00 &
  11.11 &
  0.00 &
  11.11 &
  0.00 &
  11.11 &
  0.00 &
  11.11 &
  0.00 &
  11.11 &
  0.00 &
  11.11 &
  0.00 \\ \hline
node2vec &
  3.52 &
  14.81 &
  3.35 &
  6.52 &
  5.74 &
  11.63 &
  6.21 &
  13.56 &
  3.38 &
  0.61 &
  - &
  - &
  - &
  - &
  - &
  - \\
attri2vec &
  3.93 &
  0.00 &
  4.91 &
  0.55 &
  5.69 &
  0.00 &
  6.66 &
  0.00 &
  6.5 &
  0.00 &
  - &
  - &
  - &
  - &
  - &
  - \\
GraphSAGE &
  3.36 &
  0.00 &
  2.22 &
  0.00 &
  2.06 &
  0.00 &
  1.85 &
  0.00 &
  1.72 &
  0.00 &
  - &
  - &
  - &
  - &
  - &
  - \\ \hline
pcv &
  5.60 &
  0.00 &
  2.42 &
  0.00 &
  2.00 &
  0.00 &
  1.79 &
  0.00 &
  1.70 &
  0.00 &
  1.58 &
  0.00 &
  1.62 &
  0.00 &
  1.60 &
  0.00 \\
Fraudar &
  - &
  15.64 &
  - &
  5.21 &
  - &
  3.15 &
  - &
  2.26 &
  - &
  1.76 &
  - &
  1.44 &
  - &
  1.22 &
  - &
  1.06 \\ \hline
\method &
  \textbf{11.39} &
  \textbf{64.97} &
  \textbf{13.19} &
  \textbf{86.72} &
  \textbf{13.33} &
  \textbf{97.66} &
  \textbf{13.44} &
  \textbf{99.53} &
  \textbf{13.27} &
  \textbf{100.00} &
  11.00 &
  \textbf{75.00} &
  10.72 &
  \textbf{75.00} &
  \textbf{13.41} &
  \textbf{100.00} \\
  \hline \hline 
\end{tabular}
\caption{\method consistently and significantly outperforms baselines in terms of Quality and F1 score on synthetic graphs.}
\label{tab:perform}
 \vspace{-10pt}
\end{table*}

\begin{figure*}[t]
  \centering
    \includegraphics[height=.4\linewidth]{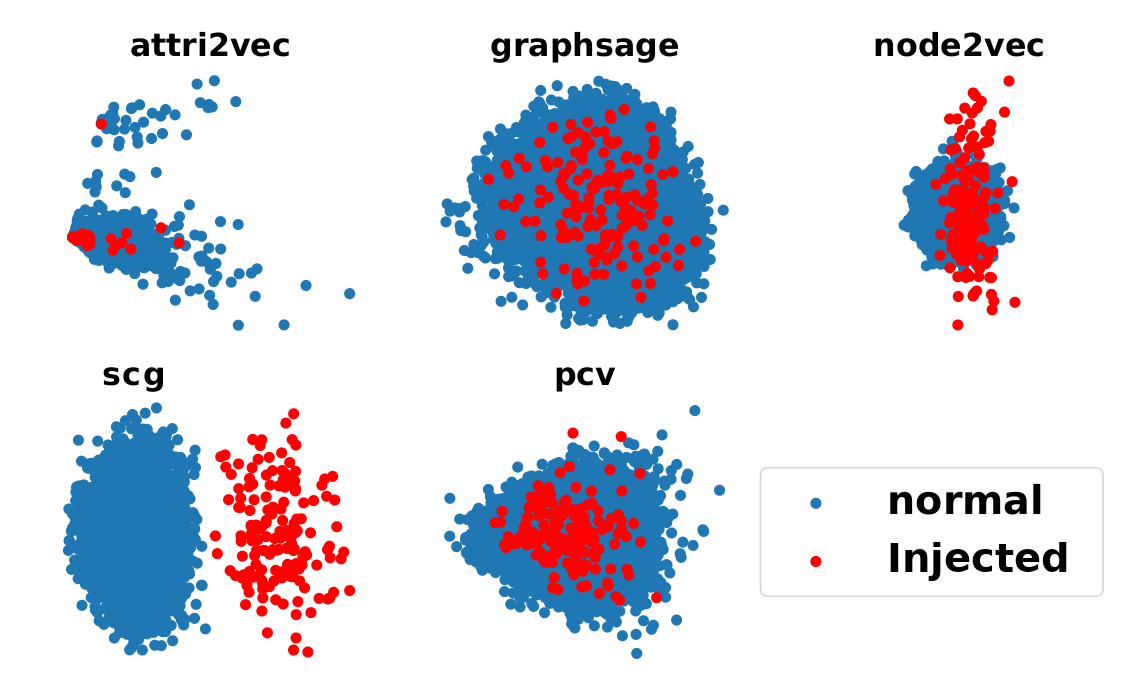}
    \vspace{-10pt}
    \caption{ \methodmap provides better separation for normal versus coordinated nodes.}%
    \label{fig:emb_cluster}%
\end{figure*}

\begin{figure*}[t]
  \centering
    \includegraphics[height=.4\linewidth]{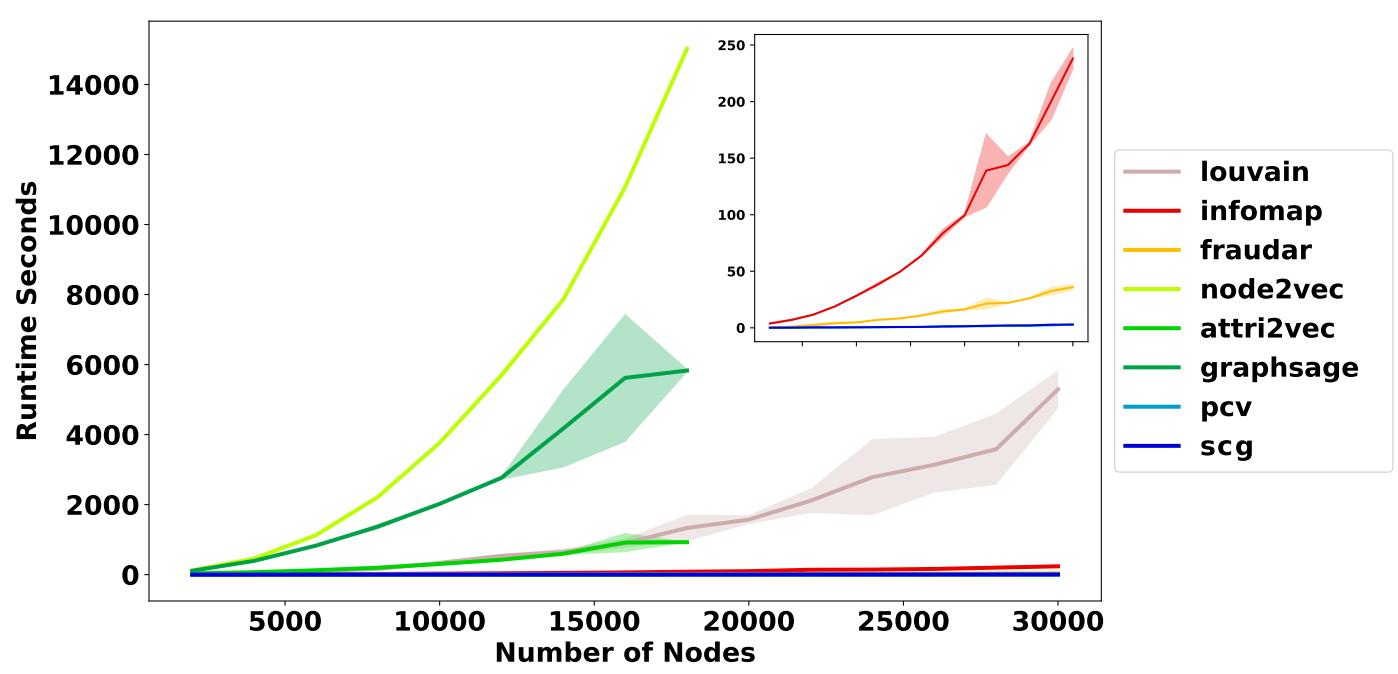}
        \vspace{-10pt}
    \caption{\methodmap is significantly faster than most baselines: more than 10,000 times faster than node2vec when graph size is 18,000. The inset plot shows the same comparison focused on the scalable algorithms.}%
    \label{fig:perform}%
\end{figure*}

Figure \ref{fig:emb_cluster} illustrates that principal components of \methodmap embeddings for normal versus \dense nodes are better separated compared to the other embeddings methods. This is an example embedding on synthetic graph of size 12,000.   Table \ref{tab:perform} reports the full results for all the baselines and settings. We can see that  \method outperforms baselines significantly, especially when the \dense groups only occupy a small fraction of the graph (0.5\%). This indicates that the general community detection or clustering methods are not appropiate for this setting as they are designed with different assumptions, e.g. balanced clusters. The pcv baseline which is specifically designed for detecting tiny clusters fails as it is not able to incorporate the connections and only operates on one mode of the data, \ie user contents.  Furthermore, as shown in Figure \ref{fig:perform}, the runtime of \methodmap is more than 10,000 times faster than some baseline. We can show that \method scales linearly with the number of nonzero entries in $\mathbf{A}$ and $\mathbf{X}$ given some assumptions. An in depth discussion on the time complexity of \method is provided in the supplementary materials.

\subsection{Results on Real-World Data}

\paragraph{\textbf{Data collection:}}
 
Since April 2019, we started collecting tweets related to the 2019 Canadian federal election through the Twitter streaming API filtered by a seed hashtag set based on significant political events in Canada (list of the hashtags used and details are provided in the supplementary materials). We collected sampled tweets between April and October 2019 and developed custom scraping pipeline to scrape all followers for Twitter users who used these hashtags. For each user, we tracked \textit{all} hashtags usage in his or her sampled tweets and created an attribute vector where each entry is the frequency of using a specific hashtag. For cross validation, we also tracked whether users been suspended between April and October, and collected their Botometer\cite{davis2016botornot} score from a commonly used API. This API measures the extent to which a Twitter account exhibits similarity to the known characteristics of social bots based on user-generated meta-data, activities, and content, without structural information about his or her follower network. For more details, please refer to the supplementary materials.

\paragraph{\textbf{Data Representation and Preprocessing:}}
 We filter out users who do not have any followers or followees, and obtain a directed attributed graph $G$ that has $n = 69,709$ nodes, $|\mathcal{E}| = 3,480,145$ edges and $d=1,329,385$ unique hashtags as node attributes.  Let $J$ denote set of all hashtags in our data ($|J| = d$), and $I$ denote the set of all users ($|I| = n$).  We create adjacency matrix $\mathbf{A}\in\{0,1\}^{n\times n}$ from user followership and attribute matrix $\mathbf{X}\in \mathbb{N}^{n\times d}$ from user hashtag usage. In the following sections, we consistently use $n$ to denote the number of users and $d$ the number of attributes. We apply doubly-normalized TF-IDF to give more significance to uncommon hashtags, because entries of $\mathbf{X}$ are highly skewed:
\begin{equation}
    \mathbf{X}^{*}_{ij} = \frac{n}{\sum_{i^\prime \in I} \mathbf{X}^b_{i^\prime j}}\frac{0.5 + 0.5 \mathbf{X}_{ij}}{\max_{j^\prime \in J}\mathbf{X}_{ij^\prime}}
\end{equation}

where $\mathbf{X}^b = \llbracket \mathbf{X} > 0 \rrbracket$ is a binarized attribute matrix.

\begin{figure}[h!]
  \centering
    \includegraphics[width=1\linewidth,trim={50cm 0 0 0},clip]{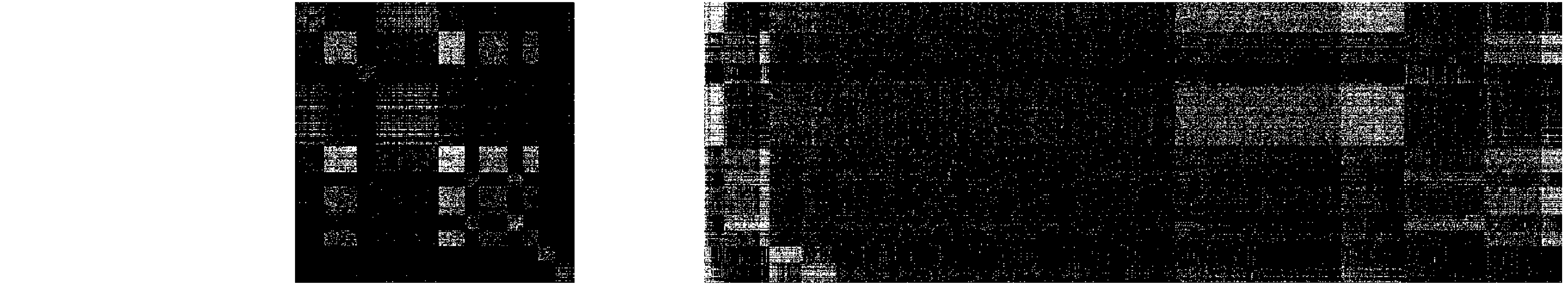}
    \caption{\methodmap finds \dense groups of users exhibiting block-diagonal structure in both the adjacency (left) and attribute matrix (right) on the twitter data.}%
    \label{fig:blocks_side}%
    \vspace{-4px}
\end{figure}

\paragraph{\textbf{Results Overview}}
A total of 13 \dense groups are detected by \method in our collected data. We visualize them in Figure \ref{fig:blocks_side}, which show a clear block structure for both $\mathbf{A}$ and $\mathbf{X}$ on indices induced by these groups. This indicates the ability for \methodmap and \methodflag to discover tightly connected user groups, each engaging with similar sets of hashtags.

\begin{figure*}
  \includegraphics[width=0.9\linewidth]{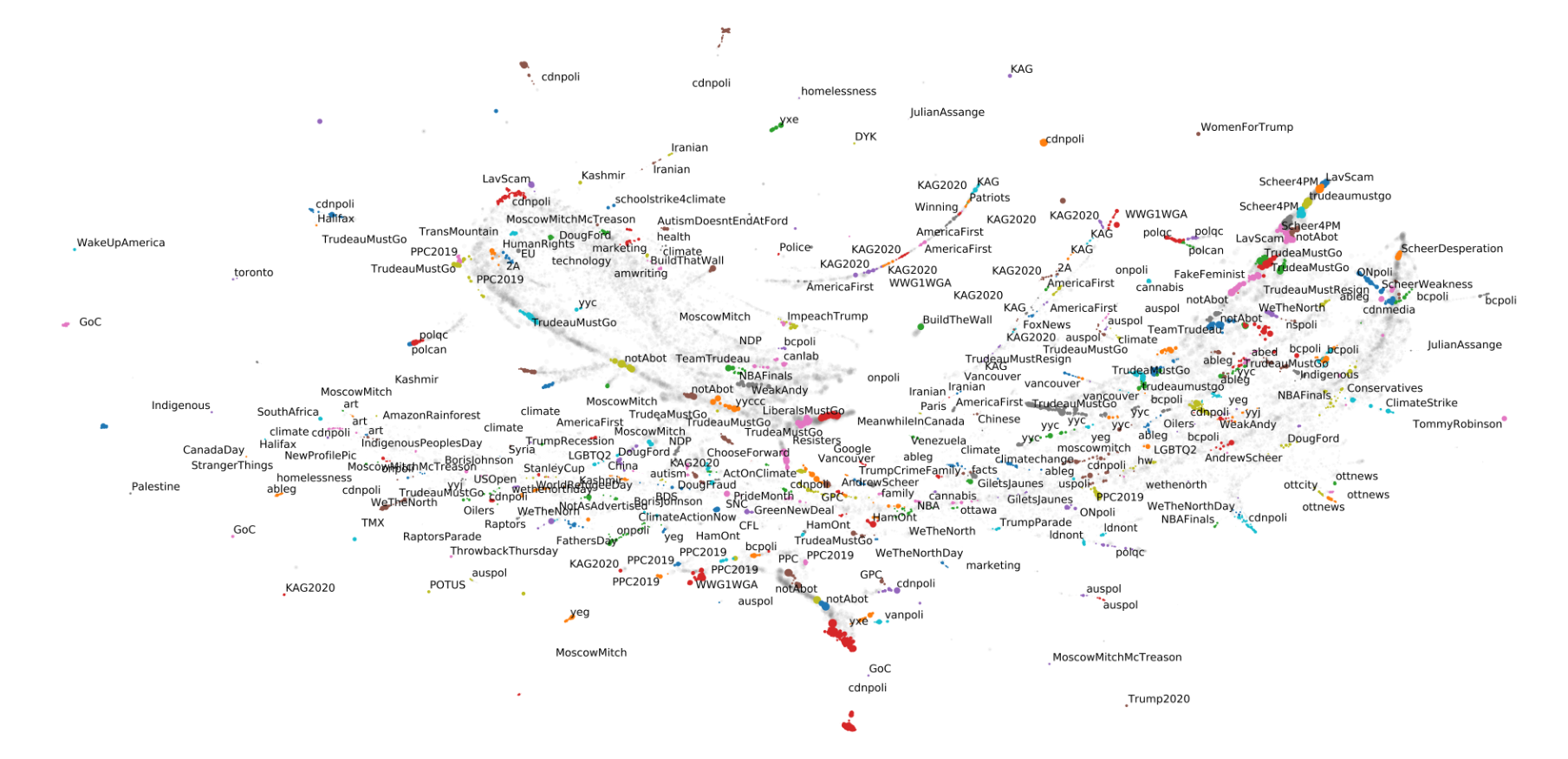}
  \caption{
  \methodmap puts users of the same political creed (related group creeds) close together. Here nodes are the individual users, size of each node corresponds to its individual engagement in the Canadian politics. Nodes are colored the same if they belong to the same cluster.  }
  \label{fig:micro}
  \vspace{-10pt}
\end{figure*}

\paragraph{\textbf{Comparing with baselines}}
We compare \method with Fraudar and pcv, which are the only baseline methods that scale to our data size given our time and hardware constraints. Since no ground truth of \dense groups is available for real-world data, we compare the suspension index and bot influence index of detected \dense groups as a proxy; which are defined below.  
Given $\mathbf{s} \in \{0,1\}^n$ to where $\mathbf{s}_{i}=1$ if user $i$ has been suspended between April and October, and $0$ otherwise, we define \textit{Suspension Index} $f_S$ of a set of user accounts $I_c$ to measure the concentration of suspended accounts in this set relative to the background, \ie: 
\begin{equation}
    f_S(I_c) = \frac{\sum_{i\in I_c}\mathbf{s}_{i}/|I_c|}{\sum_{i\in I}\mathbf{s}_{i}/|I|}
\end{equation}
Given $\mathbf{b} \in \mathbb{R}^n$ containing collected Botometer scores and $\mathbf{f} \in \mathbb{Z}^n$ containing number of followers for users in our dataset, we define \textit{Bot Influence Index} $f_B$ of a set of user accounts $I_c$ to measure their average level of estimated bot influence, \ie: 
\begin{equation}
    f_B(I_c) = \frac{\sum_{i \in I_c} \mathbf{b_i} \log(1 + \mathbf{f}_i)}{|I_c|}
\end{equation}

\begin{figure*}[t]
    \centering
  \includegraphics[width=.87\linewidth]{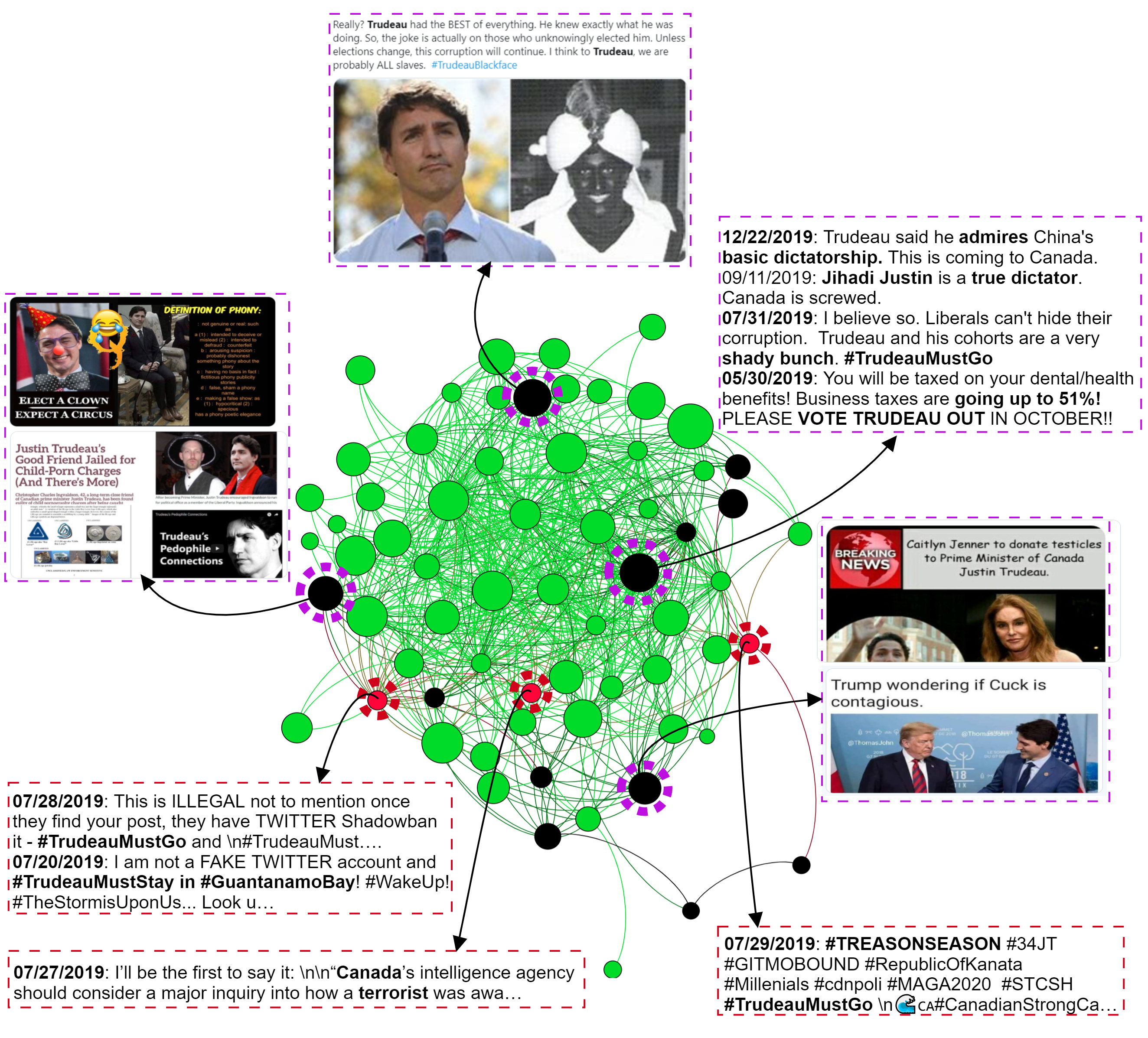}
\caption{Verification with external indicators: \methodmap and \methodflag detect an intriguing \dense group: 3 suspended users and multiple other unsuspended users simultaneously tweet politically one-sided (anti-Trudeau) and potentially offensive content.}
  \label{fig:offensive}
\end{figure*}

As shown in Table \ref{tab:real_table}, all methods perform better than uniformly sampling a set of users to be \dense, but \method is the clear winner. It detected \dense nodes that are over four times likely to be suspended than a random sample and has the highest bot influence index, which is directly related to our definition of \dense groups - set of users that boost their influence in an inauthentic fashion. Although both metrics are not designed from ground-truth knowledge of existing \dense groups, they show that \method finds interesting groups for further study, some of which we investigated in Figure \ref{fig:offensive}. Note in the figure that such high concentration of accounts that contain suspended users, posting politically one-sided (anti-Trudeau), and potentially offensive content right before the Canadian election in 2019 is intriguing.

\begin{table}[]
\centering
\begin{tabular}{c|c|c}
\hline\hline
        & Suspension Index & Bot Influence Index  \\ \hline
Fraudar & 1.297            & 1.645          \\ \hline
SPG     & \textbf{4.472}            & \textbf{1.905}       \\ \hline
pcv     & 1.625            &  1.890       \\ \hline
Random  & 1                & 0.918       \\ \hline\hline
\end{tabular}
\caption{\method detects users in \dense groups that have the highest suspension index and bot influence index.}
\label{tab:real_table}
\end{table}

\paragraph{\textbf{Discussions and Main Observations}}

Figure \ref{fig:micro} visualizes \methodmap node embeddings for users in our dataset  using UMAP \cite{mcinnes2018umap}. The sizes of the points in the figure correspond to their individual engagement in discussions around Canadian politics (Equation \ref{eq:eng}). Background nodes, those that reside in the largest cluster are plotted as grey with a lighter shade. Overlayed on each colored cluster of users is the group creed created by \methodsig. 

More specifically we define the significance of hashtag $j\in J$ denoted by $f_S$, as the mean doubly-normalized TF-IDF value across all users, \ie:
\begin{equation}
    f_S(j) = \frac{\sum_{i \in I} \mathbf{X}^*_{ij}}{n}
\end{equation}
We set 1,000 hashtags with the highest significance be the set of \textit{Significant Hashtags} $J_S$. The overlap of this set and our seed hashtag set (and their variants by changing the case of letters) gives the set of \textit{Significant Canadian Hashtags}, which we denote by $J_C$. 
We also define \textit{Individual Engagement} - each user's engagement with Canadian politics, denoted by $f_e$, as the ratio of (at-least-once) usage of hashtags in $J_C$ by that user, \ie:
\begin{equation} \label{eq:eng}
    \forall i \in I :  f_e(i) = \frac{\sum_{j \in J_C} X^b_{ij}}{|J_C|}
\end{equation}

We observe that\textit{ \methodmap embeds groups with similar group creeds close to each other, thus forming an informative map of Twitter}: top middle occupied by American conservative groups indicated by \#KAG; the center by international groups signified by \#Chinese, \#Iranian, \#Paris; top right by pro-Scheer (\#Scheer4PM) and anti-Trudeau (\#TrudeauMustGo) groups; the middle right by anti-Scheer (\#ScheerWeakness) groups; the middle left by climate activist groups, evidenced by \#climate and \#AmazonRainforest. 

The adjacency matrix with block-diagonal structure induced by the detected \dense groups is visualized in Figure \ref{fig:blockssus}, where we observe siloed groups as well as interacting ones, which are likely American conservative groups. Another observation is that the potential American conservative group signatured by \#WWG1WGA (Where We Go One, We Go All) which contains suspended users interacts with two smaller groups with the hashtag signatures of \#LavScam and \#Scheer4PM, which are likely Canadian anti-Trudeau and pro-Scheer groups. \textit{This interaction could be considered a potential foreign involvement on the Canadian 2019 Election}, which is discovered independently by other researchers after our study \cite{link4,rheault2020investigating}. Studying the impact/influence of these groups is one of our planned future studies. 

Figure \ref{fig:mesotop} illustrates the detected \dense groups, plotted as red, and non-\dense clusters that are not the background are plotted as colored points. The sizes of these points are proportional to their cluster engagement (Equation \ref{eq:enggroup}). We can see \textit{that the \dense groups are highly engaged with Canadian politics}, evidenced by their node sizes, and are close to each other in the embedding space. 
Specifically, we define \textit{Cluster Engagement} with Canadian politics for a set of users, $I_c$, as  their scaled average individual engagements with Canadian politics, \ie: 
\begin{equation} \label{eq:enggroup}
    f_E(I_c) = log(|I_c|)\frac{\sum_{i\in I_c}f_e(i)}{|I_c|}
\end{equation}

We have observed that within these \dense groups, the empirical likelihood of being \textit{suspended} between April and October is over \textit{four times more likely} compared to a random sample. Many users in these \dense groups are highly similar to those suspended accounts. We observe that\textit{ the content posted by these groups are mostly offensive}. In Figure \ref{fig:offensive} for example, in the large connected component in one of our detected \dense group, we identified several accounts (colored black) that generated politically one-sided and potentially offensive content similar to suspended accounts (colored red): some sampled content from these accounts are appended to the figure. While \methodflag spot these users who are behaving similarly to the suspended users, these accounts were still active at the time of our analysis. 


Looking at the  the group creed (signature hashtag) for each group on the Twitter maps in Figure \ref{fig:micro} and \ref{fig:mesotop} discovered by \methodsig, we get a concise characterization of the results and explains the complex structure through which these groups are engaged in Canadian politics. 
Furthermore, group creeds for \dense groups highly overlap with clusters that exhibit the highest ratio of suspended users, including \#Iranian, \#KAG2020, \#notAbot, \#TrudeauMustGo, and \#Scheer4PM. This makes \method a useful tool for spotting suspicious messages on social platforms that could have been manipulated by \dense groups. We also verify that two of \textit{these hashtags discovered by \methodsig and \methodflag (\#notAbot, \#TrudeauMustGo) are later confirmed to be linked to \textit{misinformation campaigns}} \cite{Orr19,emma19}. These two hashtags have so far been the primarily used hashtags against the 2019 Canadian election, and both have been detected before mainstream media coverage. This makes \methodsig a powerful tool to assist in detecting trending misinformation campaigns before they make a significant mark.

\methodmeso quantifies the strength of the connection between all pairs of clusters, and thus enables the study of their potential influence. 
In Figure \ref{fig:mesotop}, the link between two clusters is plotted with line width proportional to their \methodmeso interaction; those that are connected to the detected 13 \dense groups are colored red, and other links are plotted as green.
We observe from Figure \ref{fig:mesotop} that two sets of clusters have observable interactions (manifested as lines among points) among them. They are respectively represented by two sets of group creeds: (1) \#KAG (Keep America Great), \$AmericaFirst, \#WWG1WGA (Where We Go One, We Go All) and their variants which are related to American conservative politics; and (2) \#Scheer4PM, \#TrudeauMustGo, \#LavScam and their variants which are related to Canadian election politics.  Future studies will focus on the expanding this group-level study of detected \dense clusters.

A less concerning but still interesting observation is that \methodflag identifies one out of four groups signatured by \#Iranian, where two out of the four groups exhibit the highest suspension index. However, no significant connections are going outside of these three groups to other parts of the graph. Inspection of the users' tweets in these clusters reveals that the accounts in these groups are primarily concerned with immigration issues and are mostly created in February 2019, right before the passing of Bill 21, a Bill in Quebec that sets out a framework for values test for skilled workers, which impacts immigration. The observed strong connection within a set of groups but weak or no connection to other parts of a graph could be a sign of a failed amplifying strategy.

\section{Conclusions}

In this paper, we presented \method, which discovers, characterizes, and explains \dense groups in social media platforms. \method is

{\bf Holistic}: \method jointly models users' connection and content, hence it can provide the bird's eye view of the activities on social platforms. 

{\bf Effective}: \method performs significantly better than baselines on synthetic data with planted tiny clusters, and can discover intriguing \dense groups in real-world data.
 
{\bf Characterization}: \method provides a concise characterization of who (\dense groups) does what (propagating group creeds) to whom (normal groups) on large complex social networks.

{\bf  Scalable}: \method scales linearly with the number of edges of the graph with reasonable assumptions.

\bibliographystyle{plain}
\bibliography{ref.bib}

\begin{thebibliography}{10}

\bibitem{badawy2018analyzing}
Adam Badawy, Emilio Ferrara, and Kristina Lerman.
\newblock Analyzing the digital traces of political manipulation: The 2016
  russian interference twitter campaign.
\newblock In {\em 2018 IEEE/ACM International Conference on Advances in Social
  Networks Analysis and Mining (ASONAM)}, pages 258--265. IEEE, 2018.

\bibitem{blondel2008fast}
Vincent~D Blondel, Jean-Loup Guillaume, Renaud Lambiotte, and Etienne Lefebvre.
\newblock Fast unfolding of communities in large networks.
\newblock {\em Journal of statistical mechanics: theory and experiment},
  2008(10):P10008, 2008.

\bibitem{bovet2019influence}
Alexandre Bovet and Hern{\'a}n~A Makse.
\newblock Influence of fake news in twitter during the 2016 us presidential
  election.
\newblock {\em Nature communications}, 10(1):7, 2019.

\bibitem{butler2018integrating}
Andrew Butler, Paul Hoffman, Peter Smibert, Efthymia Papalexi, and Rahul
  Satija.
\newblock Integrating single-cell transcriptomic data across different
  conditions, technologies, and species.
\newblock {\em Nature biotechnology}, 36(5):411, 2018.

\bibitem{campello2013density}
Ricardo~JGB Campello, Davoud Moulavi, and J{\"o}rg Sander.
\newblock Density-based clustering based on hierarchical density estimates.
\newblock In {\em Pacific-Asia conference on knowledge discovery and data
  mining}, pages 160--172. Springer, 2013.

\bibitem{charikar2000greedy}
Moses Charikar.
\newblock Greedy approximation algorithms for finding dense components in a
  graph.
\newblock In {\em International Workshop on Approximation Algorithms for
  Combinatorial Optimization}, pages 84--95. Springer, 2000.

\bibitem{chavoshi2016identifying}
Nikan Chavoshi, Hossein Hamooni, and Abdullah Mueen.
\newblock Identifying correlated bots in twitter.
\newblock In {\em International Conference on Social Informatics}, pages
  14--21. Springer, 2016.

\bibitem{davis2016botornot}
Clayton~Allen Davis, Onur Varol, Emilio Ferrara, Alessandro Flammini, and
  Filippo Menczer.
\newblock Botornot: A system to evaluate social bots.
\newblock In {\em Proceedings of the 25th international conference companion on
  world wide web}, pages 273--274, 2016.

\bibitem{deshpande2018contextual}
Yash Deshpande, Subhabrata Sen, Andrea Montanari, and Elchanan Mossel.
\newblock Contextual stochastic block models.
\newblock In {\em Advances in Neural Information Processing Systems}, pages
  8581--8593, 2018.

\bibitem{fortunato2007resolution}
Santo Fortunato and Marc Barthelemy.
\newblock Resolution limit in community detection.
\newblock {\em Proceedings of the national academy of sciences}, 104(1):36--41,
  2007.

\bibitem{FbInauth}
Nathaniel Gleicher.
\newblock {\em How We Respond to Inauthentic Behavior on Our Platforms: Policy
  Update}, 2019 (accessed January 29, 2020).

\bibitem{mitra}
Nathaniel Gleicher.
\newblock {\em A simple algorithm for clustering mixtures of discrete
  distributions}, accessed February 29, 2020.

\bibitem{grover2016node2vec}
Aditya Grover and Jure Leskovec.
\newblock node2vec: Scalable feature learning for networks.
\newblock In {\em Proceedings of the 22nd ACM SIGKDD international conference
  on Knowledge discovery and data mining}, pages 855--864. ACM, 2016.

\bibitem{haghighat2016discriminant}
Mohammad Haghighat, Mohamed Abdel-Mottaleb, and Wadee Alhalabi.
\newblock Discriminant correlation analysis: Real-time feature level fusion for
  multimodal biometric recognition.
\newblock {\em IEEE Transactions on Information Forensics and Security},
  11(9):1984--1996, 2016.

\bibitem{hamilton2017inductive}
Will Hamilton, Zhitao Ying, and Jure Leskovec.
\newblock Inductive representation learning on large graphs.
\newblock In {\em Advances in Neural Information Processing Systems}, pages
  1024--1034, 2017.

\bibitem{DBLP:conf/kdd/HooiSBSSF16}
Bryan Hooi, Hyun~Ah Song, Alex Beutel, Neil Shah, Kijung Shin, and Christos
  Faloutsos.
\newblock {FRAUDAR:} bounding graph fraud in the face of camouflage.
\newblock In {\em {KDD}}, pages 895--904. {ACM}, 2016.

\bibitem{kelley2012defining}
Stephen Kelley, Mark Goldberg, Malik Magdon-Ismail, Konstantin Mertsalov, and
  Al~Wallace.
\newblock Defining and discovering communities in social networks.
\newblock In {\em Handbook of Optimization in Complex Networks}, pages
  139--168. Springer, 2012.

\bibitem{link2}
CIFAR Krista~Davidson.
\newblock {\em AI research detects online trolls in Canadian election}, 2019
  (accessed June 8, 2020).

\bibitem{link5}
CIFAR Krista~Davidson.
\newblock {\em CIFAR funds nine high-risk, high-reward AI projects}, 2020
  (accessed June 8, 2020).

\bibitem{kumar1999trawling}
Ravi Kumar, Prabhakar Raghavan, Sridhar Rajagopalan, and Andrew Tomkins.
\newblock Trawling the web for emerging cyber-communities.
\newblock {\em Computer networks}, 31(11-16):1481--1493, 1999.

\bibitem{kumar2017army}
Srijan Kumar, Justin Cheng, Jure Leskovec, and VS~Subrahmanian.
\newblock An army of me: Sockpuppets in online discussion communities.
\newblock In {\em Proceedings of the 26th International Conference on World
  Wide Web}, pages 857--866. International World Wide Web Conferences Steering
  Committee, 2017.

\bibitem{lee2010survey}
Victor~E Lee, Ning Ruan, Ruoming Jin, and Charu Aggarwal.
\newblock A survey of algorithms for dense subgraph discovery.
\newblock In {\em Managing and Mining Graph Data}. Springer, 2010.

\bibitem{lim2015convex}
Shiau~Hong Lim, Yudong Chen, and Huan Xu.
\newblock A convex optimization framework for bi-clustering.
\newblock In {\em International Conference on Machine Learning}, pages
  1679--1688, 2015.

\bibitem{marwick2017media}
Alice Marwick and Rebecca Lewis.
\newblock Media manipulation and disinformation online.
\newblock {\em New York: Data \& Society Research Institute}, 2017.

\bibitem{mcinnes2018umap}
Leland McInnes, John Healy, and James Melville.
\newblock Umap: Uniform manifold approximation and projection for dimension
  reduction.
\newblock {\em arXiv preprint arXiv:1802.03426}, 2018.

\bibitem{emma19}
Emma McIntosh.
\newblock A fake justin trudeau sex scandal went viral. canada's
  election-integrity law can't stop it.
\newblock {\em News, Politics, Canada's National Observer}, 2019.

\bibitem{mitchell_gottfried_kiley_matsa_2014}
Amy Mitchell, Jeffrey Gottfried, Jocelyn Kiley, and Katerina~Eva Matsa.
\newblock Political polarization and media habits.
\newblock {\em Pew Research Center}, Oct 2014.

\bibitem{mueller2019report}
Robert~S Mueller and Man With~A. Cat.
\newblock {\em Report on the investigation into Russian interference in the
  2016 presidential election}, volume~1.
\newblock US Department of Justice Washington, DC, 2019.

\bibitem{NIPS2018_7643}
Stefan Neumann.
\newblock Bipartite stochastic block models with tiny clusters.
\newblock In S.~Bengio, H.~Wallach, H.~Larochelle, K.~Grauman, N.~Cesa-Bianchi,
  and R.~Garnett, editors, {\em Advances in Neural Information Processing
  Systems 31}, pages 3867--3877. Curran Associates, Inc., 2018.

\bibitem{Orr19}
Caroline Orr.
\newblock A new wave of disinformation emerges with anti-trudeau hashtag.
\newblock {\em Election Integrity Reporting Project, Canada's National
  Observer}, 2019.

\bibitem{policy2018update}
Twitter~Public Policy.
\newblock Update on twitter’s review of the 2016 us election.
\newblock {\em Retrieved April}, 15:2018, 2018.

\bibitem{link1}
Business~Insider Rachel~Sandler.
\newblock {\em Twitter CEO Jack Dorsey reportedly shared at least 17 tweets
  from a Russian troll}, 2018 (accessed June 8, 2020).

\bibitem{link3}
LA~PRESSE RAPHAËL~PIRRO.
\newblock {\em Les trolls américains s’invitent dans la campagne}, 2019
  (accessed June 8, 2020).

\bibitem{regneri2007finding}
Michaela Regneri.
\newblock Finding all cliques of an undirected graph.
\newblock In {\em Seminar current trends in IE WS jun}, 2007.

\bibitem{rheault2020investigating}
Ludovic Rheault and Andreea Musulan.
\newblock Investigating the role of social bots during the 2019 canadian
  election.
\newblock {\em Available at SSRN 3547763}, 2020.

\bibitem{link4}
CBC~News Roberto~Rocha.
\newblock {\em Researchers found evidence of Twitter troll activity in the last
  week of the federal election}, 2019 (accessed June 8, 2020).

\bibitem{rosvall2008maps}
Martin Rosvall and Carl~T Bergstrom.
\newblock Maps of random walks on complex networks reveal community structure.
\newblock {\em Proceedings of the National Academy of Sciences},
  105(4):1118--1123, 2008.

\bibitem{TwitterElect}
Yoel Roth.
\newblock {\em Information operations on Twitter: principles, process, and
  disclosure}, 2019 (accessed January 29, 2020).

\bibitem{satopaa2011finding}
Ville Satopaa, Jeannie Albrecht, David Irwin, and Barath Raghavan.
\newblock Finding a" kneedle" in a haystack: Detecting knee points in system
  behavior.
\newblock In {\em 2011 31st international conference on distributed computing
  systems workshops}, pages 166--171. IEEE, 2011.

\bibitem{sculley2010web}
David Sculley.
\newblock Web-scale k-means clustering.
\newblock In {\em Proceedings of the 19th international conference on World
  wide web}, pages 1177--1178, 2010.

\bibitem{shin2016corescope}
Kijung Shin, Tina Eliassi-Rad, and Christos Faloutsos.
\newblock Corescope: graph mining using k-core analysis—patterns, anomalies
  and algorithms.
\newblock In {\em 2016 IEEE 16th International Conference on Data Mining
  (ICDM)}, pages 469--478. IEEE, 2016.

\bibitem{shin2016m}
Kijung Shin, Bryan Hooi, and Christos Faloutsos.
\newblock M-zoom: Fast dense-block detection in tensors with quality
  guarantees.
\newblock In {\em Joint European Conference on Machine Learning and Knowledge
  Discovery in Databases}, pages 264--280. Springer, 2016.

\bibitem{shin2017d}
Kijung Shin, Bryan Hooi, Jisu Kim, and Christos Faloutsos.
\newblock D-cube: Dense-block detection in terabyte-scale tensors.
\newblock In {\em Proceedings of the Tenth ACM International Conference on Web
  Search and Data Mining}, pages 681--689. ACM, 2017.

\bibitem{shin2017densealert}
Kijung Shin, Bryan Hooi, Jisu Kim, and Christos Faloutsos.
\newblock Densealert: Incremental dense-subtensor detection in tensor streams.
\newblock In {\em Proceedings of the 23rd ACM SIGKDD International Conference
  on Knowledge Discovery and Data Mining}, pages 1057--1066, 2017.

\bibitem{shu2019defend}
Kai Shu, Limeng Cui, Suhang Wang, Dongwon Lee, and Huan Liu.
\newblock defend: Explainable fake news detection.
\newblock In {\em Proceedings of the 25th ACM SIGKDD International Conference
  on Knowledge Discovery \& Data Mining}, pages 395--405, 2019.

\bibitem{shu2017fake}
Kai Shu, Amy Sliva, Suhang Wang, Jiliang Tang, and Huan Liu.
\newblock Fake news detection on social media: A data mining perspective.
\newblock {\em ACM SIGKDD Explorations Newsletter}, 19(1):22--36, 2017.

\bibitem{starbird2019disinformation}
Kate Starbird.
\newblock Disinformation’s spread: bots, trolls and all of us.
\newblock {\em Nature}, 571(7766):449, 2019.

\bibitem{stewart2018examining}
Leo~G Stewart, Ahmer Arif, and Kate Starbird.
\newblock Examining trolls and polarization with a retweet network.
\newblock In {\em Proceedings of WSDM workshop on Misinformation and
  Misbehavior Mining on the Web (MIS2)}, 2018.

\bibitem{wang2019sgp}
Junhao Wang, Sacha Levy, Ren Wang, Aayushi Kulshrestha, and Reihaneh Rabbany.
\newblock Sgp: Spotting groups polluting the online political discourse.
\newblock {\em arXiv preprint arXiv:1910.07130}, 2019.

\bibitem{wilson2018assembling}
Tom Wilson, Kaitlyn Zhou, and Kate Starbird.
\newblock Assembling strategic narratives: Information operations as
  collaborative work within an online community.
\newblock {\em Proceedings of the ACM on Human-Computer Interaction},
  2(CSCW):183, 2018.

\bibitem{xu2014jointly}
Jiaming Xu, Rui Wu, Kai Zhu, Bruce Hajek, Rayadurgam Srikant, and Lei Ying.
\newblock Jointly clustering rows and columns of binary matrices: Algorithms
  and trade-offs.
\newblock In {\em The 2014 ACM international conference on Measurement and
  modeling of computer systems}, pages 29--41, 2014.

\bibitem{yang2011detecting}
Tianbao Yang, Yun Chi, Shenghuo Zhu, Yihong Gong, and Rong Jin.
\newblock Detecting communities and their evolutions in dynamic social
  networks—a bayesian approach.
\newblock {\em Machine learning}, 82(2):157--189, 2011.

\bibitem{zhang2019attributed}
Daokun Zhang, Jie Yin, Xingquan Zhu, and Chengqi Zhang.
\newblock Attributed network embedding via subspace discovery.
\newblock {\em Data Mining and Knowledge Discovery}, 33(6), 2019.

\bibitem{zhang2017hidden}
Si~Zhang, Dawei Zhou, Mehmet~Yigit Yildirim, Scott Alcorn, Jingrui He, Hasan
  Davulcu, and Hanghang Tong.
\newblock Hidden: hierarchical dense subgraph detection with application to
  financial fraud detection.
\newblock In {\em Proceedings of the 2017 SIAM International Conference on Data
  Mining}, pages 570--578. SIAM, 2017.

\bibitem{zhou2018fake}
Xinyi Zhou and Reza Zafarani.
\newblock Fake news: A survey of research, detection methods, and
  opportunities.
\newblock {\em arXiv preprint arXiv:1812.00315}, 2018.

\bibitem{zhou2019fake}
Xinyi Zhou, Reza Zafarani, Kai Shu, and Huan Liu.
\newblock Fake news: Fundamental theories, detection strategies and challenges.
\newblock In {\em Proceedings of the Twelfth ACM International Conference on
  Web Search and Data Mining}, pages 836--837, 2019.

\end{thebibliography}

\pagebreak
\section{Appendix}

In this section, we provide details of problem, methods, experiments and algorithm scalability. We also created an interactive visual demo based on \method to enable the interpretability of our work. Source code for synthetic experiment and URL for interactive visual demo can be retrieved at \url{https://sites.google.com/view/spg-exp}. We run all experiments on Acer Predator Triton 700 PT715-51-732Q Notebook with Intel Core i7-7700HQ (2.80 GHz), 32GB DDR4, and NVIDIA GeForce GTX 1080 8GB. The budget for collecting individual tweet history that spans multiple months and writing custom scraping pipeline to scrape the full follower network of the large number of users is 5,000 Canadian dollars.

\subsection{Problem Motivation}
We define Coordinated Group to be (1) a small set of densely connected social media accounts that aim to increase their influence through (2) following each other and broadcasting a similar set of messages. 

\paragraph{\textbf{(1) Small network:}}

Quoting from the report of U.S. Department of Justice's investigation into Russian interference in the 2016 U.S. Presidential election: "\textbf{Dozens of IRA employees} were responsible for operating accounts and personas on different U.S. social media platforms; \textbf{A number of} IRA employees assigned to the Translator Department served as Twitter specialists; IRA specialists operated certain Twitter accounts to \textbf{create individual U.S. personas}" \cite{mueller2019report}, evidence suggests that the size of the \dense group is small and operationalized, and the total number of employees operating the social media accounts is limited by hiring capacity of the underlying organization.

\paragraph{\textbf{(2) Following each other and broadcasting messages:}}

According to investigation lead by the Special Counsel for the United States Department of Justice, Russian "Active Measures" Social Media Campaign executed by Internet Research Agency, LLC (IRA) was capable of reaching millions of U.S. citizens through their social media accounts on Facebook, Instagram, Tumblr, YouTube, and Twitter, by the end of the 2016 U.S. election \cite{mueller2019report}. More specifically, IRA created inauthentic social media accounts operated by a small team of employees as well as automated bots starting as early as 2014, in the names of U.S. citizens, fictitious U.S. organizations and grassroots groups, in order to garner followers and influence in online discourse and broadcast messages with hidden political agenda. Employee-operated IRA social media accounts attracted massive followers: "United Muslims of America" Facebook group had over 300,000 followers; $@$jenn\_abrams - a Twitter account claiming to be a Virginian Trump supporter had over 70,000 followers. Bot-operated network of accounts also gained considerable influence during the election (approximately 1.4 million people on Twitter) \cite{mueller2019report}.

\subsection{Details of Problem Formulation}

Formally, we model the users as a binary attributed graph $G(\mathbf{A}, \mathbf{X})$, where $\mathbf{A} \in \{0,1\}^{n\times n}$ is the adjacency matrix that encodes connections among users, and $\mathbf{X} \in \{0,1\}^{n\times d}$ is the binary attribute matrix that encodes connections between users and content. We assume both $\mathbf{A}$ and $\mathbf{X}$ can be modelled by Bernoulli random variables in some space set, whose distribution can be specified through latent groups on the rows and columns. Let node affiliation matrix $\mathbf{Z} \in \{0, 1\}^{n \times g}: \sum_{k=1}^d \mathbf{Z}_{ik} = 1$ be a binary matrix that encodes hard partition of $n$ nodes into $g$ non-overlapping latent node groups, and attribute affiliation matrix $\mathbf{W} \in \{0, 1\}^{d \times m}$ be a binary matrix that encodes hard partition of $d$ attributes into $m$ possibly overlapping latent attribute groups. Let $\mathbf{P}\in[0,1]^{g\times g}$ store the means of a grid of Bernoulli random variables where $\mathbf{P}_{kk^\prime} = p(\mathbf{A}_{ii^\prime} = 1 \ | \ \mathbf{Z}_{ik} \mathbf{Z}_{i^\prime k^\prime} = 1 )$, and similarly $\mathbf{Q} \in [0,1]^{g \times m}$ where $\mathbf{Q}_{kl} = p(\mathbf{X}_{ij} = 1 \ | \ \mathbf{Z}_{ik} \mathbf{W}_{jl} = 1 )$. $\mathbf{P}_{kk^\prime}$ represents the probability of having an edge between two nodes belonging to latent node groups $k$ and $k^\prime$, respectively, while $\mathbf{Q}_{kl}$ represents the probability that a user from latent node group $k$ has attribute from latent attribute group $l$. Assume $\mathbf{A}$ and $\mathbf{X}$ are conditionally independent given $\mathbf{Z}, \mathbf{W}$ and $\boldsymbol{\theta} = \{\mathbf{P}, \mathbf{Q}\}$, thus:
\begin{equation}
    \begin{split}
        p(\mathbf{A}, \mathbf{X} | \mathbf{Z}, \mathbf{W}, \boldsymbol{\theta}) &= p(\mathbf{A}| \mathbf{Z}, \mathbf{W}, \boldsymbol{\theta}) p( \mathbf{X} | \mathbf{Z}, \mathbf{W}, \boldsymbol{\theta})= \\
   \prod_{ii^\prime kk^\prime} f(\mathbf{A}_{ii^\prime}; \ & \mathbf{P}_{kk^\prime})^{ \mathbf{Z}_{ik}\mathbf{Z}_{i^\prime k^\prime}} \ \prod_{ijkl} f(\mathbf{X}_{ij}; \ \mathbf{Q}_{kl})^{ \mathbf{Z}_{ik}\mathbf{W}_{kl}} \\
   \text{where } f(a;b) &= a^b(1-a)^{1-b}
    \end{split}
\end{equation}

Given such a latent variable model we define a \dense group to be a latent node group $k\in \{1,\cdots,g\}$ that is small, well connected within itself, and share at least one small latent attribute group $l\in\{1,\cdots,m\}$ that nodes in group $k$ have with high probability. Formally, we defined as:

\begin{definition}[block \dense group] \label{def:pollutegroup}
Given threshold parameters for group-induced subgraph edge probabilities $p^*\in[0,1]$, $q^*\in[0,1]$, and group size bounds \footnote{$s_h $ is in the order of $ O(n^\epsilon)$ and $t_h$ is in the order of $ O(d^\epsilon)$ for some $\epsilon>0$ to confine \dense groups to be tiny clusters} $s_h \geq s_l\in\mathbb{N}, t_h \geq t_l\in\mathbb{N}$, a latent node group $k\in\{1,\cdots g\}$ of $s$ nodes is called a \textbf{\dense group} with shared latent attribute group $l\in \{1,\cdots,m\} $ of $t$ attributes if it satisfies the following conditions:

\begin{itemize}
\item \textbf{Size Condition} (groups are small):
$$s_l\leq s \leq s_h \text{ and } t_l\leq t \leq t_h$$
\item \textbf{Edge Probability Condition} (group is well connected and attribute is likely):
$$\mathbf{P}_{kk}\geq p^* \text{ and }\mathbf{Q}_{kl}\geq q^*.$$
\end{itemize}
 
\end{definition}

\subsection{Details of Methods}
\label{apd:dom}
We model the activity on social networks (Twitter, Facebook) as a binary attributed graph $G(\mathbf{A},\mathbf{X})$, which potentially contains \dense groups as defined in Definition \ref{def:pollutegroup}. In cases where it is more optimal to have non-binary attributes, we relax the binary constraint. We design \method to be a modular framework that consists of four components: \methodmap \textbf{maps out} large scale activity on social media by jointly embedding user connections and the content they post; \methodflag \textbf{detects} groups of users which are posting similar content and are also densely connected to each other, a common indicator of misinformation campaigns, which we call \dense groups; \methodsig  \textbf{characterizes} the engagement of the \dense groups and finds their group creed; \methodmeso  \textbf{explains} how different \dense groups engage with each other and the rest of the population. 

Specifically, \textbf{\methodmap} first \textbf{jointly embeds} nodes and their attributes of $G(\mathbf{A},\mathbf{X})$ to low-dimensional Euclidean vectors $\mathbf{Z}$ by preserving one-hop neighborhood structural similarity and attribute information, which we view as a form of data fusion, where multiple sources of data are integrated to produce more consistent, accurate and useful information than using a single source \cite{haghighat2016discriminant}. We design four variants of \methodmap, which we call \methodmap-original (referred to as \methodmap is the main paper), \methodmap-augment, \methodmap-lanigiro, \methodmap-tnemgua (abbreviated as $f_o, f_a, f_l, f_t$), based on two fundamental building blocks: low-rank approximation of matrix denoted by projection operator $\Pi_K$, where $K$ specifies the resulting dimension after projection; and message passing on graph denoted by operator $M(\cdot \ ; \mathbf{A})$, where $\mathbf{A}$ encodes adjacency matrix of the graph. The variants differ in the order with which to apply $\Pi$ and $M$, as well as whether to apply augmentation $T$ on $\mathbf{A}$:
\begin{equation}
    \begin{split}
        f_o(\mathbf{A},\mathbf{X}) &= \mathbf{Z}_o= M(\Pi_K(\mathbf{X});\mathbf{A})\\
        f_a(\mathbf{A},\mathbf{X}) &= \mathbf{Z}_a=M(\Pi_K(\mathbf{X}); T(\mathbf{A}))\\
        f_l(\mathbf{A},\mathbf{X}) &= \mathbf{Z}_l=\Pi_K(M(\mathbf{X};\mathbf{A}))\\
        f_t(\mathbf{A},\mathbf{X}) &= \mathbf{Z}_t=\Pi_K(M(\mathbf{X};T(\mathbf{A})))
    \end{split}
\end{equation}
Arbitrary $\Pi_K, M, T$ can be plugged in to produce a mapping from $G(\mathbf{A},\mathbf{X})$ to $\mathbf{Z}$. In this work, we chose $\Pi_K$ to project input matrix on its first $K$ left singular vectors, and chose $M$ to use summation aggregator for graph message passing, and chose $T$ to augment input adjacency matrix by adding self-loops to all nodes. For example, $f_t$ materializes as $\Pi_K(\mathbf{A}\mathbf{X} + \mathbf{X})$. $\Pi_K$ can be efficiently implemented by augmented Lanczos bidiagonalization algorithm \cite{butler2018integrating}. 

The functional form of \methodmap, especially $f_l$ is motivated by (1) interpreting $\widetilde{\mathbf{D}} = M(\mathbf{X},\mathbf{A})$ as the extension of degree matrix to binary attributed graph, where $\widetilde{\mathbf{D}}_{ij}$ represents number of neighbors with attribute $j$ that node $i$ has; (2) interpreting rows of $\widetilde{\mathbf{D}}$ to be drawn from a mixture of discrete probability distributions on $\mathbb{N}^d$. It has been proven that for a matrix of size $n\times d$ whose rows are vectors sampled from a mixture of $d$-dimensional discrete distributions or continuous distributions with subgaussian tails, under certain conditions, applying low-rank projection and then $l_2^2$ clustering can partition these row vectors to their respective generating distributions exactly with high probability, even if the partition size is tiny - in the order of $O(n^\epsilon), \epsilon > 0$ \cite{mitra}. Therefore, our \methodmap followed by \methodflag can be viewed as the exact recovery of tiny user clusters with similarly attributed neighbors with high probability, echoing the intuition of \dense groups in Definition \ref{def:pollutegroup}. Furthermore, since such procedure works for both discrete and continuous data, attribute matrix $\mathbf{X}$ can be relaxed to be non-binary.

\textbf{\methodflag} detects \dense groups in $G$ by first constructing clusters through minibatch KMeans - a scalable centroid-based clustering optimized by stochastic gradient descent \cite{sculley2010web}. In principle, however, other clustering algorithms can be used. For example, when the ground truth number of clusters is not known, density-based clustering can be used. Definition \ref{def:pollutegroup} states that for a cluster to be a \dense group, it needs to satisfy edge probability condition and size condition for both $\mathbf{A}$ and $\mathbf{X}$. Note, however, to check both conditions on $\mathbf{X}$ requires inferring the underlying generative parameters of the binary attributed graph, which is not the approach taken in our work, and we leave it for future work. Thus, we only check edge probability condition and size condition for $\mathbf{A}$ to flag a cluster to be a \dense group, and check only edge probability condition for $\mathbf{A}$ in cases where the size threshold is unknown.

\textbf{\methodsig} creates a concise interpretable group signature for each cluster generated by \methodflag which we call \textbf{group creed}, by defining a metric $\phi$ over the set of attributes $J$ given nodes $I_c$ of a cluster, to rank the informativeness of each attribute in terms of uniquely describing node cluster at $I_c$.
\begin{equation}
    \forall j \in J: \ \phi(j; I_c) = \frac{\sum_{i \in I_c}\mathbf{X}_{ij}}{\sum_{j^\prime \in J}\sum_{i \in I_c} \mathbf{X}_{i j^\prime}} - \frac{\sum_{i \in I}\mathbf{X}_{ij}}{\sum_{j^\prime\in J}\sum_{i\in I} \mathbf{X_{ij^\prime}}}
\end{equation}
$\phi(j ;I_c)$ shows the local versus global discrepancy of usage frequency for attribute $j$. Therefore, $argmax_{j\in J}\phi(j;I_c)$  is the group signature (group creed) for cluster $I_c$, and it gives a concise summarization for constructed clusters from \methodflag, both \dense and normal ones. 

\textbf{\methodmeso} defines a symmetrical pairwise metric $\psi$ between two sets of nodes $I_c$ and $I_p$:
\begin{equation} \label{eq:psi}
    \psi(I_c, I_p) = \frac{\sum_{i\in I_c}\sum_{i^\prime \in I_p} \mathbf{A}_{ii^\prime}}{|I_c||I_p|}
\end{equation}
$\psi(I_c, I_p)$ captures the strength of the interaction between nodes in $I_c$ and $I_p$, thus providing a meso-level view of the data.

The four components of \method above make it a powerful tool for joint \dense group detection and data summarization on attributed graphs. Next, we demonstrate its effectiveness, scalability, and interpretability with synthetic experiments and real-world data applications.

\subsection{Details of Experiments on Synthetic Data}

\paragraph{\textbf{Parameter settings:}} 
For threshold parameters, we set $p^*=q^*=0.01, s_l=t_l = 10, s_h=t_h =80$, essentially limiting \dense group-induced subgraph to be appropriately sized with edge probability higher than 0.01. For generative parameters, we set $\mathbf{P}\in\{0.025,0.015,0.01, 0.005\}^{9\times 9}$ where $\mathbf{P}_{kk} = 0.025, \mathbf{P}_{kk^\prime} = 0.015,\mathbf{P}_{k9}= \mathbf{P}_{9k} = 0.01,\mathbf{P}_{99} = 0.005$ for $k,k^\prime \in [1\dots 8],k\neq k^\prime$. We set $\mathbf{Q}\in\{0.025, 0.005\}^{9\times 9}$ where $\mathbf{Q}_{ll} = 0.025$ for $l\in[1\dots 8]$ and other entries of $\mathbf{Q}$ to be 0.005. For node affiliation matrix $\mathbf{Z}\in\{0,1\}^{n\times 9}$, we set $\mathbf{Z}_{i1} = 1$ for $i\in[1\dots 20]$, $\mathbf{Z}_{i2} = 1$ for $i\in[21\dots 40]$, $\dots$, $\mathbf{Z}_{i9} = 1$ for $i\in[161\dots n]$, and other entries of $\mathbf{Z}$ to be 0. This encodes 8 small latent groups that induce subgraphs on $\mathbf{A}$ with higher edge probabilities which correspond to \dense groups, and 1 large latent group with low edge probability, which corresponds to normal users. Note that each \dense group also has denser connections with other \dense groups and normal users than the connections among normal users themselves. For attribute affiliation matrix $\mathbf{W}\in\{0,1\}^{d\times 9}$, we set $\mathbf{W}_{j9} = 1 \ \forall j\in[1\dots d]$, and for each $l\in[1\dots 8]$, we create index set $J_l: |J_l| = 40$ by uniformly random sampling without replacement from $[1 \dots d]$, and set $\mathbf{W}_{jl} = 1$ for $j \in J_l$. This encodes 8 small possibly overlapping latent groups and 1 large latent group that contains all attributes, meaning all attributes (hashtags) are equally likely to be used by non-\dense nodes. 

\paragraph{\textbf{Evaluation:}} 
When generating the data, we assign a label for each node where labels 1 to 8 denotes which coordinated group a node belongs to, and if a node does not belong to any coordinated group,
we assign label 0. We also assign a binary label to each node that
indicates whether the node belongs to any coordinated group or not. 

\paragraph{\textbf{Baselines:}}

Infomap, Louvain, node2vec only run on adjacency matrix $\mathbf{A}$ and cannot utilize attribute matrix $\mathbf{X}$; pcv only runs on bipartite graph, thus only $\mathbf{X}$ and not $\mathbf{A}$; we run Fraudar on $\mathbf{A}$; our method \method, attri2vec, and GraphSAGE run on both $\mathbf{A}$ and $\mathbf{X}$. For algorithms that return partition of both nodes and attributes (pcv), we evaluate partition accuracy by the Quality score between ground truth and inferred partition, and evaluate classification accuracy of nodes being in \dense group or not by assigning estimated binary labels to nodes through sequentially checking (1) the edge probability condition for $\mathbf{A}$; (2) the size condition for $\mathbf{A}$; (3) both conditions for $\mathbf{X}$, and report the best F1 score from the three. For algorithms that return partition of nodes (Infomap, Louvain, \method), we follow the procedure above but do not check for (3) For algorithms that return embedding vectors of nodes (node2vec, attri2vec, GraphSAGE), we first apply minibatch KMeans clustering with number of cluster equal to 9 and create node labels, and then proceed as those mentioned above. For an algorithm that only returns the anomalous set of nodes (Fraudar), we only report the F1 score. Any algorithm that already has a centroid-based clustering step involved (pcv), we replace it with minibatch KMeans for fairness of comparison. For any algorithm that has an embedding component, if the chosen implementation has a default setting that is consistently used or if the original publication explicitly recommends a particular setting, we set the embedding dimension accordingly. Otherwise, we set it to 10 for fairness of comparison.

\subsection{Details of Experiments on Real-World Data}

\paragraph{\textbf{Data collection:}}

The seed set of hashtags are in Table \ref{tab:hashtags}. Using Google Cloud virtual machine instance, we started to use Twitter streaming API to collect real-time sampled tweets that contain these hashtags since April. In the meantime, our custom scraping pipeline scanned the real-time tweets and initiated massively parallel jobs to scrape the full follower network (not easily accessible through Twitter API) for each user in that dataset, as well as large-size sample of individual tweet history that spans multiple months. Till October, the size of scraped data was around 2-3 terabytes before filtering. The scraping, cleaning and processing of this dataset have been both labor and capital intensive. 

\begin{table}[h!]
\centering
\footnotesize
\begin{tabular}{|l|l|}
\hline
\#cdnpoli & \#canpoli \\ \hline
\#cpc & \#SenCA \\ \hline
\#cdnleft & \#pttory \\ \hline
\#ptbloc & \#gpc \\ \hline
\#crtc & \#goc \\ \hline
\#BlackFaceTrudeau & \#TrudeauMustResign \\ \hline
\#BlackFace & \#BrownFace \\ \hline
\#ScheerLies & \#elexn43 \\ \hline
\#NotasAdvertised & \#TrudeauTheHyprocrite \\ \hline
\#ptlib & \#lpc \\ \hline
\#ndp & \#lavscam \\ \hline
\#ptndp & \#ptgreen \\ \hline
\#cdnsen & \#cpac \\ \hline
\#CdesCom & \#TrudeauBlackFace \\ \hline
\#BrownFaceTrudeau & \#TrudeauWorstPM \\ \hline
\#Scheer & \#Andysresume \\ \hline
\#elxn43 & \#elxn19 \\ \hline
\end{tabular}
\caption{Hashtags used for crawling the data which are related to 2019 Canadian Federal Election.}
\label{tab:hashtags}
\end{table}

\paragraph{\textbf{Applying \methodmap}}
We chose \methodmap-original for its efficiency. Compared with synthetic data, some of the users in the twitter data do not have any followers, and simple message-passing is not optimal. Thus we relax \methodmap by message-passing on both $\mathbf{A}$ and $\mathbf{A^T}$ and then concatenate to form final node embedding $\widetilde{\mathbf{H}}$. We set the projected dimension $K$ of $\Pi_K$ to be 100, thus resulting in $\Tilde{\mathbf{H}}\in\mathbb{R}^{n\times 200}$:
\begin{equation}
\begin{split}
    \widetilde{\mathbf{H}} &= \big(f_o(\mathbf{A}, \mathbf{X}^*),f_o(\mathbf{A}^T, \mathbf{X}^*)\big) \\
\end{split}
\end{equation}

\begin{figure}
  \centering
    \includegraphics[height=0.5\linewidth]{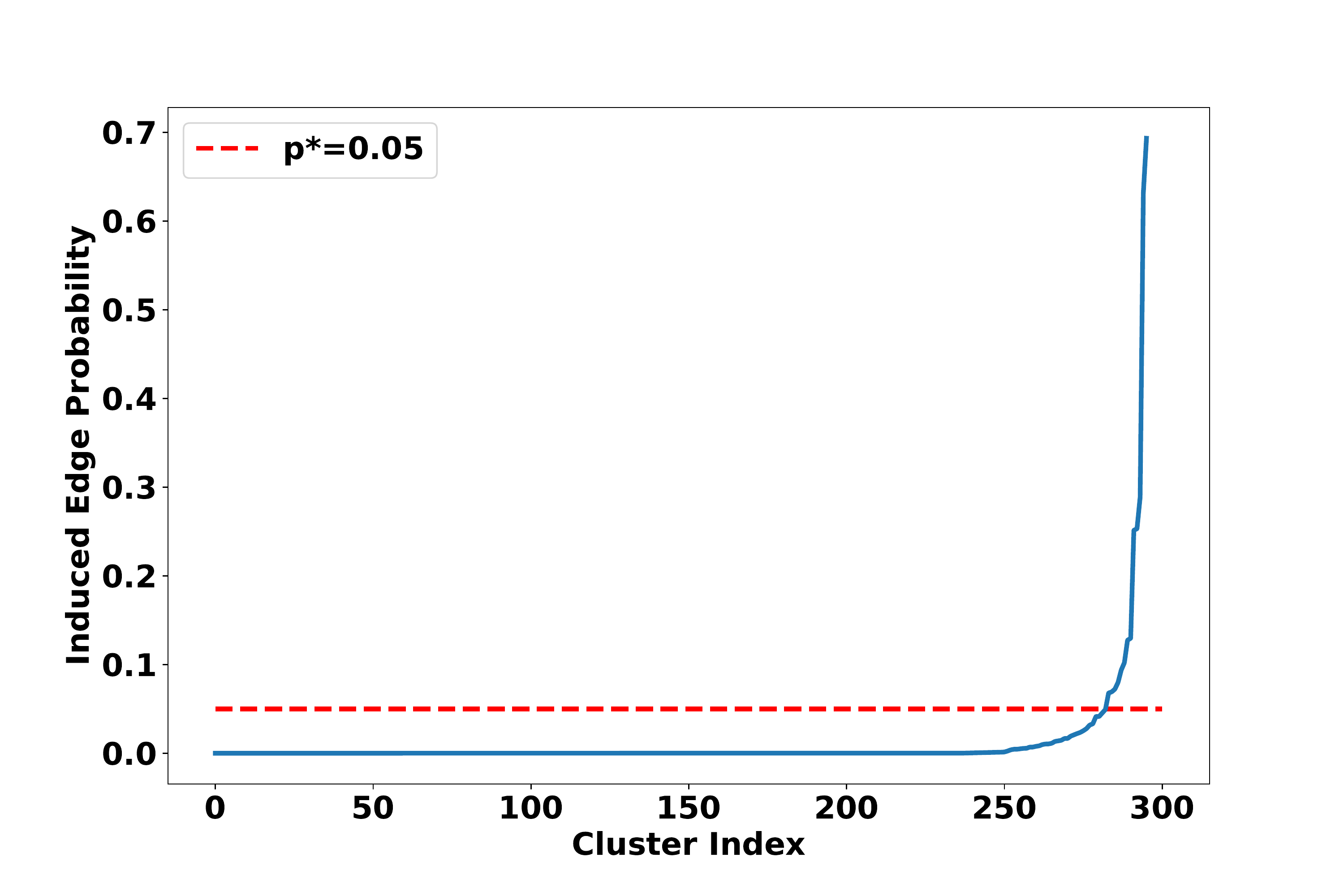}
   
    \caption{Given cluster index sorted by induced subgraph edge probability on $\mathbf{A}$, we apply elbow method to identify $p^*=0.05$ for \dense groups}%
    \label{fig:elbow}%
\end{figure}

\paragraph{\textbf{Applying \methodflag}}
Compared with synthetic experiments, both ground truth number of clusters as well as threshold parameters for \dense groups defined in Definition \ref{def:pollutegroup} are not known. Therefore we applied a clustering algorithm that does not require the knowledge of the number of clusters and then applied elbow method \cite{satopaa2011finding} to determine $p^*$, the minimum edge probability of cluster-induced subgraph on $\mathbf{A}$. We selected HDBSCAN \cite{campello2013density}, a density-based clustering algorithm that does not require the knowledge of number of clusters, and set $p^* = 0.05$ through elbow heuristics as in Figure \ref{fig:elbow}. 

\paragraph{\textbf{Applying \methodsig}}

Naive application of \methodsig on the set of all hashtags $J$ would result in uninformative signatures, especially since $|J|$ is large. Therefore, we apply \methodsig to the set of Significant Hashtags $J_S$. Thus, given the set of all users $I$, the group signature (creed) for the cluster of users $I_c$ is $argmax_{j\in J_S}\phi(j; I_c)$ where

\begin{equation}
    \forall j \in J_S: \ \phi(j; I_c) = \frac{\sum_{i \in I_c}\mathbf{X}_{ij}}{\sum_{j^\prime \in J_S}\sum_{i \in I_c} \mathbf{X}_{i j^\prime}} - \frac{\sum_{i \in I}\mathbf{X}_{ij}}{\sum_{j^\prime\in J_S}\sum_{i\in I} \mathbf{X_{ij^\prime}}}
\end{equation}

\paragraph{\textbf{Applying Fraudar and pcv}}

We run Fraudar on the adjacency matrix $\mathbf{A}$ of follower network. We run pcv on the transformed attribute matrix $\mathbf{X}^*$, and convert non-binary labels returned from pcv to binary labels in the same way as in synthetic experiments. We set embedding dimension of pcv to be 100, clustering module of pcv to be HDBSCAN, and $p^* = 0.05$, same as \method.

\subsection{\bf \scale \ of \method}

Let $|\cdot|$ be the number of nonzero entries in a sparse matrix. We assume input matrices $\mathbf{A}\in\{0,1\}^{n\times n},\mathbf{X}\in\mathbb{R}^{n\times d}$ are sparse and their number of nonzero entries are in the same order: $O(|\mathbf{X}|)\approx O(|\mathbf{A}|)$, and assume the projection dimension $K$ in \methodmap, number of inferred clusters $m$ in \methodflag, batch-size $B$ and number of steps $T$ for any stochastic iterative algorithms used in \method are much smaller than $|\mathbf{A}|$, i.e. $K\ll|\mathbf{A}|, m\ll|\mathbf{A}|, B\ll|\mathbf{A}|, T\ll|\mathbf{A}|$. We analyze the time complexity of \method composed of (1) \methodmap-original, (2) \methodflag clustering based on minibatch KMeans, (3) \methodflag thresholding, \methodsig and \methodmeso implemented with hash map through a constant number of passes of all nonzero entries of sparse matrices. The time complexity for \method can be calculated as:

\begin{itemize}
    \item \textbf{\methodmap-original}: First applying augmented Lanczos bidiagonalization algorithm \cite{butler2018integrating} to calculate $\Pi_K(\mathbf{X})$ takes $O(T|\mathbf{X}|K + K^3 + c)\approx O(|X|)\approx O(|A|)$. Next, for message passing operation $M(\Pi_K(\mathbf{X}),\mathbf{A})$ where $\Pi_K(\mathbf{X})\in\mathbf{R}^{n\times K}$ is dense and $\mathbf{A}$ is sparse, the time complexity is $O(|A|K)\approx O(|A|)$ for standard implementation using Sparse Basic Linear Algebra Subprograms (BLAS) Library.
    \item \textbf{\methodflag clustering}: Applying minibatch KMeans to \methodmap-original node embeddings $\mathbf{Z}\in\mathbb{R}^{n\times K}$ take $O(KBmT) < O(|\mathbf{A}|)$.
    \item \textbf{Other components:} Since they are implemented to take constant number of passes through all nonzero entries of input sparse matrices, the time complexity is $O(|\mathbf{A}|)$.
\end{itemize}

Therefore, the resulting time complexity of \method is $O(|A|)$, thus scaling linearly with the number of edges in the graph.

\subsection{\method Interactive Visualization Dashboard}
As one of the future works of this paper, we are building an interactive tool to share our findings with collaborators in political science and journalism. The demo URL can be accessed at \url{https://sites.google.com/view/spg-exp}.

\end{document}